\documentclass[11pt,twocolumn]{article}
\usepackage{latexsym}
\normalbaselines
\begin{document}
\twocolumn[
\centerline{{\Large\bf On The New Symmetries in Electrodynamics and 
Quantum Theory}}
\medskip
\centerline{{\bf Gennadii KOTELNIKOV}}

\centerline{{\bf Russian Research Center Kurchatov Institute}}

\centerline{{\bf Kurchatov Sq., 1, Moscow, 123182, Russia}}
\bigskip]

\centerline{{\bf ABSTRACT}}
\medskip
\begin{sloppypar}\noindent
The generalized definition of symmetry is formulated.  Application  of
this   definition   for  symmetric  analysis  of  theoretical  physics
equations is considered. The version of electrodynamics is constructed
permitting  the  faster-than-light  motions  of  particles  with  real
masses. Some elements  of  physical  interpretation  of  the  proposed
theory are presented.
\end{sloppypar}
\medskip

\centerline{\bf 1.INTRODUCTION}
\medskip

\begin{sloppypar}
\noindent
The study  on the symmetric properties of theoretical and mathematical
physics equations is a  standard  and  important  task  of  scientific
research.  It is connected with the fact that the symmetric properties
of theoretical physics equations contain  fundamental  information  on
the properties of the world around.  As an illustration we shall dwell
on three examples.
\end{sloppypar}
\begin{sloppypar}
Let us  consider  the  Newton dynamics equation (1687)
\begin{equation}\label{eq.1}
m{\bf a}={\bf F},
\end{equation}
where $m$  is  the mass of a particle,  ${\bf a}$ is the acceleration,
${\bf F}$ is the force acting on a particle. It is well-known that the
equation is   invariant   with   respect  to  the  Galilei  space-time
transformations
\end{sloppypar}
\begin{equation}\label{eq.2}
x'=x-Vt, \ y'=y, \ z'=z, \ t'=t,
\end{equation}
where $c'=c(1-2\beta  n_x+\beta^2)^{1/2}$ is the law of transformation
of  the  velocity  of  light  $c$  with  respect  to   transformations
(\ref{eq.2}),  $\beta=V/c$,  ${\bf  n}={\bf  c}/c$,  $(x,y,z)$ are the
spatial variables,  $t$ is the time, ${\bf V}=(V,0,0)$ is the velocity
of inertial motion of the reference frame $K'$ relative to $K$.  It is
follows from here that the space surrounding us is  3-dimensional  and
possesses the Euclid metric
\begin{equation}\label{eq.3}
ds^2=dx^2+dy^2+dz^2.   
\end{equation}
Time is universal at any point of space.  Space is empty and serves as
arena where events have been taking place. Space does not influence on
these  events.  The  3-dimensional  coordinate  space  and  time exist
independently of one another.

These concepts  were  kept  until  1873  when  Maxwell  equations were
discovered
\begin{equation}\label{eq.4}
\begin{array}{ll}
\vspace{1mm}
\displaystyle
\nabla{\rm X}{\bf E}+\frac{1}{c}\frac{\partial{\bf H}}{\partial t}=0, &
\displaystyle
\nabla{\rm X}{\bf H}-\frac{1}{c}\frac{\partial{\bf E}}{\partial t}=
\frac{4\pi}{c}\rho{\bf v},                                           \\
\displaystyle
\nabla\cdot{\bf E}=4\pi\rho, & \nabla\cdot{\bf H}=0.
\end{array}
\end{equation}
\begin{sloppypar}\noindent
(${\bf E},  \  {\bf  H}$  are  the  electric   and   magnetic   fields
respectively,  $\rho$  is the electrical charge density,  ${\bf v}$ is
the velocity of electrical charge, $c$ is the speed of light). In 1904
to  1905 in the fundamental works of Lorentz,  Einstein and Poincar\'e
it was shown that the Maxwell equations are invariant with respect  to
the Lorentz transformations
\begin{equation}\label{eq.4a}
\begin{array}{l}
\displaystyle
x'=\frac{x-Vt}{\sqrt{1-V^2/c^2}}, \
t'=\frac{t-Vx/c^2}{\sqrt{1-V^2/c^2}}, 
\end{array}
\end{equation}
where $y'=y,  \ z'=z, \ c'=c$. The consequence from the found symmetry
is a profound change in our ideas of space and  time.  Time  has  been
ceased  to  be  independent.  Space  has  become  4-dimensional and is
characterized by the pseudo-Euclid metric
\begin{equation}\label{eq.5}
ds^2=c^2dt^2-dx^2-dy^2-dz^2,  
\end{equation}
into which  the  differentials  of time and coordinates enter on equal
rights.  The constant of the speed of light plays a fundamental  role.
But  space,  as  in the case of classic electrodynamics,  is empty and
does not exert any effect on dynamics of physical processes  happening
in it.
\end{sloppypar}

\begin{sloppypar}
But in 1928 the Dirac equation was formulated
\begin{equation}\label{eq.6}
(\gamma^ap_a-mc)\Psi(x)=\frac{e}{c}\gamma^aA_a\Psi(x),
\end{equation}
where $x=(x^0,{\bf  x})$,  $x^0=ct$,  $e$  is  the  electric   charge,
$A^a=(A^0,{\bf  A})$  is  the  4-potential,  $\gamma^aA_a=\gamma^0A^0-
\gamma^1A^1-\gamma^2A^2-\gamma^3A^3$,   $\gamma^a$   are   the   Dirac
matrices. It  is  invariant  with respect to the transformation of the
special form $C$ known as the charge conjugation.  This transformation
is accompanied by the reversal of signs of electric charge, energy and
momentum.  The idea of anti-particle,  positron,  and Dirac electronic
vacuum  comes  into being.  Empty space,  vacuum,  ceased to be empty.
Vacuum is filled with electrons at the states with negative  energies.
Vacuum influences on quantum-mechanical processes going on in the real
world.
\end{sloppypar}

\begin{sloppypar}
These are the consequences of the physical interpretation of  abstract
mathematical  symmetries  in  theoretical  physics.  It  will  not  be
overstated if we relate the D'Alembert and the Shr\"odinger  equations
to such equations. They have extensive applications in experiment, are
well studied through the known methods of symmetry analysis and  form,
as well  as  the equations of Newton,  Maxwell and Dirac,  a basis for
modern physics. The natural question arises of whether these equations
contain  something  more  than  what  is  known  of  them.  It  is not
improbable  that  they  have  additional  symmetries  which   are   as
significant for physics as the ones already found.  Thus,  the purpose
of the present work is the finding of additional  symmetries  inherent
in the equations of D'Alembert, Dirac, Maxwell and Schr\"odinger.
\end{sloppypar}
\bigskip

\centerline{\bf 2.DEFINITION OF SYMMETRY}
\medskip

\begin{sloppypar}\noindent
In solving this problem, it is necessary to take into account a number
of the  rigorous  theorems:  the  13-dimensional  Schr\"odinger  group
$Sch_{13}$ is the maximal group of symmetry of Schr\"odinger  equation
in the space of classical physics (Niederer,  1972,  Hagen, 1972); the
15-dimensional conformal  group  $C_{15}$  is  the  maximal  group  of
symmetry  of  the equations of D'Alembert and Maxwell in the Minkowski
space (Pauli, 1921).
\end{sloppypar}

\begin{sloppypar} 
It is  essential  that  all  these  symmetries  relate to so named the
Lie-type symmetries in the space  of  classical  physics,  or  in  the
Minkowski  space.  It  is  follows  from here that for finding the new
symmetries it is necessary either to be beyond the space of  classical
physics  and the Minkowski space or to turn to the non-Lie generalized
types of symmetry.  It is in this way that the new results  have  been
obtained.  Let  us  turn  first  to  the task of searching for the new
symmetries on the base of the generalized understanding  of  symmetry.
\end{sloppypar}

\begin{sloppypar}
Let some linear partial differential equation $L\phi(x)=0$  be  given.
By   symmetry  of  this  equation  we  shall  understand  the  set  of
$Q$-operators satisfying the permutation relations of the $p$-th order
$[L...[L,Q]...]]_{(p-fold)}\phi(x)=0$ on the solutions $\phi(x)$. When
$p=1$,  this definition coincides with the standard  one  (Malkin  and
Man'ko, 1965; Kyriakopoulos, 1968).
\end{sloppypar}

\begin{sloppypar}
As an illustration,  let us use this definition for the one  component
field  $\phi(x)$ when $p=2$,  and the symmetry operators $Q$ belong to
some           Lie           algebra:            $[L[L,Q]]=\zeta(x)L$;
$Q=\xi^a(x)\partial_a+\eta(x)$;       $[Q_l,Q_m]=C_{lmn}Q_n$.       By
substituting the expression for the  $Q$-operator  into  the  operator
equality and by equating the coefficients with the same derivatives on
the left and on the right,  a  set  of  the  determining  differential
equations   for   the  unknown  functions  $\xi^a(x)$,  $\eta(x)$  and
$\zeta(x)$ may be obtained.  Commutation properties of  the  operators
enable  one  to  find  the  structure  constants  and  to identify the
algebra.  On the base of this algebra,  the coordinate transformations
$x'=x'(x)$   may   be   found   by   integrating   the  Lie  equations
$dx^{a'}/d\theta=\xi^a(x'),  \ {x^{a'}}_{\theta=0}=x^a$,  ($\theta$ is
the  group  parameter).  Let  us determine the law of transforming the
field function by the relation  $\phi'(x')=\Phi(x)\phi(x)$.  Here  the
weight  function  $\Phi(x)$ may be calculated from the set of engaging
equations $A\Phi(x)\phi(x)=0,  \ L\phi(x)=0$,  where  the  former  was
obtained  by  replacing  the  variables in the initial primed equation
$L\phi'(x')=0$. The compatibility of the set $ A\Phi\phi=0, \ L\phi=0$
is,  according  to  the definition,  the condition of transforming the
initial   equation   into   itself   $L\phi'(x')=0   \to    (x'=x'(x),
\phi'(x')=\Phi(x)\phi(x)) \to A\Phi(x)\phi(x)=0, \ L\phi(x)=0$ [1].
\end{sloppypar}
\medskip

\begin{sloppypar}
\centerline{\bf 3.GALILEI INVARIANCE OF}
\centerline{\bf MAXWELL EQUATIONS}
\end{sloppypar}
\medskip

\begin{sloppypar}\noindent
Let us  apply  the  algorithm to the equations chosen for analysis and
consider first the D'Alembert equation.  We start  from  the  equation
$L\phi=\Box\phi=0$  for  scalar  field and the solution in the form of
plane  waves   $\phi=e^{-i\omega(t-{\bf   n}\cdot{\bf   x}/c)}$.   The
equations  and  the  solutions  will be considered within the space of
classical  physics.  Let  us  introduce  the  Galilei  transformations
(\ref{eq.2})  and  turn  to  the  permutation relations of the Galilei
transformations generator $H=t\partial_x$ with the D'Alembert operator
$[\Box[\Box,H]]=0$.  It  can  be  seen  from  here  that  the  Galilei
transformations   for   the   D'Alembert   equation    are    symmetry
transformations of  the  type  $p=2$.  Let  us  do  the replacement of
variables  taking  into  account  the  law  of  field   transformation
$\phi'=\Phi\phi$  and  find  the weight function from the condition of
compatibility      of       the       equation       $A\Phi(x)\phi(x)=
[(\partial_0+\beta\partial_x)^2/\lambda^2-\triangle]\Phi\phi=0$   with
the equation $\Box\phi(x)=0$. The weight function takes the form
\begin{equation}\label{eq.10}
\displaystyle
\Phi(x)=e^{-i[(1-\lambda)\omega(t-{\bf n}\cdot{\bf x}/c)-\beta\omega
(n_xt-x/c)]/\lambda},
\end{equation}
where ${\bf n}={\bf c}/c$, $\lambda=c'/c$. In this case the D'Alembert
equation  transforms into itself $\Box\phi'(x')=0\to \Box\phi(x)=0$ in
accordance with the standard understanding  of  symmetry.  To  Galilei
invariance of   D'Alembert  equation  there  corresponds  the  Galilei
invariance of the light cone equation $c^2t^2-{\bf x}^2=0$  (it can be
proved by direct calculation).
\end{sloppypar}

Let us  turn now to the Maxwell equations and consider the homogeneous
equations  and  their  solutions  in  the  form  of  the  plane  waves
$({\rm{\bf  E},{\bf  H}})=({\bf  l},{\bf  m})\cdot  e^{-i\omega(t-{\bf
n}{\bf x}/c)}$, where ${\bf l}, {\bf m}$ are the polarization vectors.
Here   the   field,   as   distinct   from   the   previous  case,  is
multi-component.  Therefore,   we   shall   search   for   the   field
transformation law in the more complicated form
\begin{equation}\label{eq.11}
\begin{array}{l}
{{\rm E}'}_x=\Phi(x){\rm E}_x,                   \\
{{\rm E}'}_y=\Phi(x)k({\rm E}_y+h_{yz}{\rm H}_z),\\
{{\rm E}'}_z=\Phi(x)k({\rm E}_z+h_{zy}{\rm H}_y),\\
{{\rm H}'}_x=\Phi(x){\rm H}_x;                   \\
{{\rm H}'}_y=\Phi(x)k({\rm H}_y+e_{yz}{\rm E}_z);\\
{{\rm H}'}_z=\Phi(x)k({\rm H}_z+e_{zy}{\rm E}_y).
\end{array}
\end{equation}
\begin{sloppypar}\noindent
Here $\Phi(x)$  is  the weight function (\ref{eq.10}) corresponding to
the condition of  the  D'Alembert  equation  invariance;  $k,  e_{yz},
e_{zy},  h_{yz},  h_{zy}$  the  parameters  of  field transformations.
By replacing the variables,  taking into account the  weight  function
$\Phi(x)$,  we  find  a  set  of 8 algebraic equations for the unknown
values $k,  e_{yz}, e_{zy}, h_{yz}, h_{zy}$. The set is compatible and
has the solutions:
\end{sloppypar}
\begin{equation}\label{eq.12}
\begin{array}{l}
\displaystyle
k=+\frac{n_x(\beta-n_x)+\lambda}{1-{n_x}^2},                  \\
\displaystyle
e_{yz}=+\frac{n_x(\lambda-1)+\beta}{n_x(\beta-n_x)+\lambda},  \\
\displaystyle
h_{yz}=-\frac{n_x(\lambda-1)+\beta}{n_x(\beta-n_x)+\lambda}, 
\end{array}
\end{equation}
where $e_{yz}=-e_{zy}=h_{zy}=-h_{yz}$.    In   accordance   with   the
generalized understanding of  symmetry,  it  means  that  the  Maxwell
equations  are  invariant  with respect to the Galilei transformations
[1].  In  the  limit  of  low  velocities  the  formulae   for   field
transformations  take  the  form  known under the name of the Galilean
limit
\begin{equation}\label{eq.13}
\displaystyle
{\rm{\bf E}}'\approx{\rm{\bf E}}+\beta{\rm X}{\rm{\bf H}}, \
{\rm{\bf H}}'\approx{\rm{\bf H}}-\beta{\rm X}{\rm{\bf E}}.
\end{equation}
These relationships  form,  in  fact,  the  same  limit  both  in  the
relativistic and in the Galilei case. Here the situation is similar to
the situation concerning the  transformations of  space-time  variables
when $v\ll c$
\begin{equation}\label{eq.14}
x'=x-Vt, \ y'=y, \ z'=z, \ t'=t, \ c'=c.
\end{equation}
These transformations, being neither the Galilean nor the relativistic
transformations,  are the limiting relationships for both  the  cases.
\bigskip

\centerline{\bf 4.LORENTZ INVARIANCE OF} 
\centerline{\bf SCHR\"ODINGER EQUATION}
\medskip

\noindent
Let us    next    consider    Schr\"odinger     equation     $L_S\psi=
(i\hbar\partial_t+\hbar^2\triangle/2m_0)\psi(x)=0$.  We start from the
generators  $M_{ok}=x^0\partial_k-x^k\partial_0$,  $k=1,2,3$  of   the
Lorentz   transformations.   In  view  of  the  permutation  relations
$[L_S[L_S,M_{ok}]]=0$,   the   Lorentz   transformations    are    the
Schr\"odinger  equation  symmetry  transformations  of the type $p=2$.
Using it as the base,  we shall extend the  Shr\"odinger  equation  to
the relativistic  domain  of  motions  by introducing the relativistic
mass  $m=m_0/\sqrt{1-\beta^2}$  and  the  relativistic  energy  ${\cal
E}=mc^2$ of a particle
\begin{equation}\label{eq.15}
\Big(i\hbar\partial_t+c^2\hbar^2\frac{\triangle}{2{\cal E}}\Big)\psi^r(x)=0.
\end{equation}
The equation has two solutions:
\begin{equation}\label{eq.16}
\begin{array}{l}
\displaystyle
{\psi^r}_1(x)=e^{-i(mv^2/2\hbar)(t-2{\bf s}\cdot{\bf x}/v)}; {} \\
\displaystyle
{\psi^r}_2(x)=e^{-i(mc^2/\hbar)(t-\sqrt2{\bf s}\cdot{\bf x}/c)},
\end{array}
\end{equation}
where ${\bf s}={\bf v}/v,  {\bf x}=(x,y,z),  {\bf v}=(v_x,v_y,v_z)$ is
the  velocity  of  a  particle.  Besides,  equation  (\ref{eq.15})  is
invariant with respect to Lorentz transformations (\ref{eq.4a}) if the
weight   function  $\Psi$  from  the  relation  ${\psi^r}'=\Psi\psi^r$
satisfies the                                                 equation
$[i\hbar(\partial_t+V\partial_x)+(c^2\hbar^2(1-\beta^2)/2{\cal      E}
(1-{\bf                                                    V}\cdot{\bf
v}/c^2))[(\partial_x+\beta\partial_t/c)^2/(1-\beta^2)+
\partial_{yy}+\partial_{zz}]]\Psi\psi^r=0$,  \ ${\bf V}=(V,0,0)$.  The
weight functions  corresponding  to  the  solutions  (\ref{eq.16}) are
${\Psi^r}_1$ and ${\Psi^r}_2$,  take a rather  complicated  form,  and
will  not  be  presented  here.  The  solutions  (\ref{eq.16})  in the
nonrelativistic approximation take the form
\begin{equation}\label{eq.17}
\begin{array}{l}
\displaystyle
{\psi^r}_1(x)\to \psi_1(x)=
e^{-i(Et-{\bf p}\cdot{\bf x})/h}; {} \\
\displaystyle
{\psi^r}_2(x)\to \psi_2(x)=
e^{-i({\cal E}_0t-\sqrt2{\bf p}\cdot{\bf x}/\beta)/\hbar}, 
\end{array}
\end{equation}
\begin{sloppypar}\noindent
where $E=m_0v^2/2$,   ${\cal   E}_0=m_0c^2$,   ${\bf  p}=m_0{\bf  v}$;
$\psi_1(x)$  is  the  known   solution;   $\psi_2(x)$   is   the   new
linear-independent  solution  that  reveals itself  in  analyzing  the
Lorentz invariance in the Schr\"odinger  equation.  By  comparing  the
results  obtained  with  the  well-known  results  published,  one can
conclude that the Galilei group is the symmetry group in the sense  of
Lie  for  the  Schr\"odinger  equation  and  the symmetry group in the
generalized  sense  for  the  equations  of  D'Alembert  and  Maxwell.
Analogously,  the  Lorentz group is the symmetry group in the sense of
Lie for the equations of D'Alembert and Maxwell and the symmetry group
in the generalized sense for the Schr\"odinger equation [1].
\end{sloppypar}
\bigskip

\centerline{\bf 5.FIVE-DIMENSIONAL SPACE WITH}
\centerline{\bf NONINVARIANT VELOCITY OF LIGHT}
\medskip

\small
\begin{sloppypar}\noindent
Let us  dwell  on  examples  of the new symmetries associated with the
change-over  to  the  space  of  different  dimensionality.  We  shall
introduce 5-dimensional abstract space of events $V^5(t,{\bf x},c)$ in
which  the  velocity  of  light  $c$  plays  the  role  of  additional
independent    variable.   Next,   we   shall   consider   commutation
relationships of the generators $g_s$ of the conformal group  $C_{15}$
and  the generators $X_n=c^n[c\partial_c - t\partial_t + N(t\partial_t
+ {\bf x}\cdot\nabla)]$ of the  Virasoro  infinite  algebra  with  the
D'Alembert operator in this space
\end{sloppypar}
\begin{equation}\label{eq.18}
\begin{array}{l}
\displaystyle
[\Box,g_s]=0, \  g_s=\{c^NP_a,M_{ab}\};  \\
\displaystyle
[\Box,g_p]\phi=0, \ g_p=\{D,c^{-N}K_a\}; \\
\displaystyle
[g_s,g_p]=C_{spq}g_q;                    \\
\displaystyle
[X_n,X_m]=(m-n)X_{m+n};                  \\
\displaystyle
[X_n,g_s]=0.
\end{array}
\end{equation}
Here $g_s$ belong to a set of the operators $c^NP_a$,  $M_{ab}$,  $D$,
$c^{-N}K_a$;  $N=0,\pm1,\pm2\dots$;  $C_{spq}$  are  the  Lie  algebra
structure   constants  of  the  conformal  group;  $|m|,|n|  <\infty$.
According to the generalized definition of symmetry,  this algebra  is
the invariance algebra of the D'Alembert equation and the Maxwell free
equations  of  the  type  $p=1$.  The  algebra  may  be  brought  into
correspondence  with  the  coordinate  transformations  group with the
noninvariant velocity of light.  In the particular  case  $N=0$  these
transformations  were  considered by Romain (1963),  the author of the
present work (1970), Di Jorio (1974), Sj\"odin (1979). With $c'=c$ the
formulae  contain  the  Voigt  transformations  (1887),  Ives  (1937),
Palacios-Gordon (1957,  1962),  Dewan  (1961),  Podlaha  (1969).  With
$c'=\lambda  c$  ($\lambda$  is  the group parameter) they contain the
transformations of the author of the present work (1970),  Hsu  (1976)
and Mamaev (1990) [1].
\bigskip

\centerline{\bf 6.DISCRETE "-C" SYMMETRY}
\medskip

\noindent
We shall next turn to the new symmetry of discrete type.  As is known,
the discreet symmetries play an important role in modern physics,  for
example,  in particle physics,  quantum field theory, nuclear physics.
One can note,  as an example,  the space inversion $P({\bf  x}\to-{\bf
x})$,  the time inversion $T(t\to-t)$, the charge conjugation $C$. Let
us introduce a new discreet transformation - inversion of the velocity
of light $Q: t\to t, \ {\bf x}\to{\bf x}, \ c\to-c$. The inversion $Q$
is closed into a group that forms a direct product  with  the  Lorentz
group (it does not change the form of the Lorentz transformations) and
is the new transformation of discrete symmetry for the  equations  of
classic  and quantum electrodynamics:  the equations of D'Alembert and
Maxwell,  the  equations  for  movement  of  a  charged  particle   in
electromagnetic  field,  the  Klein-Gordon-Fock  equation,  the  Dirac
equation  and   the   Schr\"odinger   equations   (\ref{eq.15}).   The
consequence  of  $-c$ symmetry of Dirac equation is the possibility to
interpret the charge conjugation $C$ in terms of  the  $Q$-conjugation
$[C,Q]\Psi(x)=0$, where $\Psi(x)$ is the solution of Eq.  (\ref{eq.6})
[1].
\bigskip

\centerline{\bf 7.INTERNAL SYMMETRY OF}
\centerline{\bf  MAXWELL EQUATIONS}
\medskip
\begin{sloppypar}\noindent
Besides the  space-time symmetries,  in physics applications also find
so  named  internal  symmetries.  The  internal  symmetries  are   not
connected  with  the properties of space and time and characterize the
very objects to be studied.  As applied to electromagnetic field,  the
investigation  of  such  symmetries  was  started by Heaviside (1893),
Rainich  (1925)  and  Larmor  (1928).  It  was  then  carried  out  by
Markhashov  (1966),  Danilov  (1967)  and  Ibragimov (1968) within the
framework of the Lie classic algorithm.  Fushchich and Nikitin  (1978)
and  then  the  author  (1982)  of  the  present  work continued these
investigations with the help of the  non-Lie  algorithm  (proposed  by
Fushchich and Nikitin) on the base of the Fourier transformations.  We
shall use the elements of  this  algorithm,  start  from  the  Maxwell
homogeneous  equations and introduce the Fourier transformations ${\bf
E}(x^0,{\bf x})=(1/2\pi)^{3/2}\int{\bf\cal E}(x^0,{\bf p})e^{i{\bf  p}
\cdot{\bf x}}d^3p$, \ ${\bf H}(x^0,{\bf x})=(1/2\pi)^{3/2}\int{\bf\cal
H}(x^0,{\bf p})e^{i{\bf p} \cdot{\bf x}}d^3p$ for electric  ${\bf  E}$
and  magnetic ${\bf H}$ field and turn to the Maxwell equations in the
$p$-space
\begin{equation}\label{eq.18a}
\begin{array}{ll}
i{\bf p}{\rm x}{\bf\cal E}+\partial_0{\bf\cal H}=0, &
i{\bf p}{\rm x}{\bf\cal H}-\partial_0{\bf\cal E}=0, \\
{\bf p}\cdot{\bf\cal E}=0, & {\bf p}\cdot{\bf\cal H}=0.
\end{array}
\end{equation}
From the  condition  of  invariance of these equations with respect to
the transformations  of  fields  ${\bf\cal  E}\to  {\bf\cal  E}'$  and
${\bf\cal  H}\to{\bf\cal H}'$ we find a set of the sixteen $6{\rm X}6$
matrices   $Y_{LMN}(p),   \    Z_{LMN}(p)$,    \    $p=(p_1,p_2,p_3)$,
$(L,M,N)=0,1$   transforming   the   equations   (\ref{eq.18a})   into
themselves.  The matrices satisfy the permutation relations of the Lie
algebra, the Grassman algebra and super algebra $[Y,Y]=Y; \ [Y,Z]=Z; \
\{Z,Z\}=Y$.  Out of them the Lie 16-dimensional algebra is  isomorphic
to  the  Lie  algebra  of  the unitary transformations group $U(2){\rm
X}U(2){\rm X}U(2){\rm X}U(2)$,  which contains the 8-dimensional group
$U(2){\rm X}U(2)$ of Fushchich and Nikitin and the 2-dimensional group
$U(1){\rm X}U(1)$ of Danilov and Ibragimov.  The latter  includes  the
Larmor-Rainich transformations   ${\bf    E}'={\bf    E}cos\theta+{\bf
H}sin\theta,  \ {\b H}'=-{\bf E}\sin\theta+{\bf H}cos\theta$ ($\theta$
is the group parameter) and  the  discreet  Heaviside  transformations
${\bf  E}'= {\bf H},  \ {\bf H}'=-{\bf E}$ at $\theta=\pi/2$.  All the
transformations in the $x$-space,  except  for  the  Danilov-Ibragimov
transformations, are nonlocal (integral) [1].
\end{sloppypar} 
\vspace{5mm}

\centerline{\bf 8.NONLINEAR MAXWELL EQUATIONS}
\medskip

\noindent
In all the cases considered we dealt with  linear  equations.  But  in
contemporary  physics nonlinear equations has received wide acceptance
as well. Nonlinear equations are used, for example, in the catastrophe
theory,  in the theory of control,  in quantum theory.  They possess a
number of such exclusive properties as the  absence  of  superposition
principle,  the  existence  of  nonlinear  interaction of fields,  the
existence of soliton solutions.  The first work on nonlinear equations
in  electrodynamics was done by Mie in 1912 on the base of introducing
functions of field invariants into the theory. The concrete version of
nonlinear  equations  was proposed by Born and then by Born and Infeld
(1934). The nonlinear electrodynamics equations in the present work
\begin{equation}\label{eq.19}
\begin{array}{l}
\vspace{1mm}
\displaystyle
\nabla{\rm X}{\bf E}+\frac{1}{c}\frac{\partial{\bf H}}{\partial t}=0, \\
\vspace{1mm}
\displaystyle
\nabla{\rm X}{\bf H}-\frac{1}{c}\frac{\partial{\bf E}}{\partial t}=
\frac{4\pi}{c}F(I_1,I_2)\rho{\bf v},\\
\vspace{1mm}
\displaystyle
\nabla\cdot{\bf E}=4\pi F(I_1,I_2)\rho, \\
\displaystyle  
\nabla\cdot{\bf H}=0.
\end{array}
\end{equation}
differ from the  Born-Infeld  equations  and  are  of  self-sufficient
importance.   (Here   $F$   is   the   arbitrary   function   of   the
Lorentz-invariants of  the  field  $I_1=2({\bf  E}^2  -  {\bf  H}^2)$,
$I_2=({\bf E}{\bf H})^2$). With $\rho=0$ they contain the Maxwell free
equations;  with $F(I_1,I_2)=1$ they go  into  the  Maxwell  equations
(\ref{eq.4});  the  nonlinearity is caused by the presence of currents
and charges.  The equations realize the principle of self-action  with
the  result  that  the  electric  charge within the framework of these
equations partially has a field nature.  In view of  the  relativistic
invariance  the  equations  are of potential interest for physics [1].
\vspace{5mm}

\centerline{\bf 9.APPLICATION TO PHYSICS}
\smallskip

\begin{sloppypar}\noindent
Next we shall turn to the question of physical interpretation  of  the
mathematical  results,  in  particular,  from  Section {\bf 5}.  As an
example we shall consider the symmetry connected with violation of the
Special Relativity (SR) second postulate - postulate of the invariance
of the speed of  light.  The  questions  on  the  possibility  of  its
violation  and  on  the  existence  of  the  faster-than-light motions
invariably attracts  the  scientific  community's  attention.  Abraham
(1908),  Ritz (1908),  Blokhintsev (1946), Kirzhnits (1954), Terletsky
(1960),  Rapier (1962),  Feinberg (1967), Loiseau (1968), Bilaniuk and
Sudarshan  (1969),  the  author  of  the present work (1970),  Marinov
(1975),  Hsu (1976),  Recami (1982),  Logunov  (1982),  Chubykalo  and
Smirnov-Rueda (1996),  Russo (1998),  Glashow (1999) and other authors
investigated this problem from different  points  of  view.  Below  we
shall  consider the author's work [2] which starts from the invariance
of 4-interval of space-time
\begin{equation}\label{int}
ds^2=({c_0}^2+v^2)dt^2-dx^2-dy^2-dz^2,
\end{equation}
where  the invariant $c_0=3\cdot10^{10}$ cm/sec is the proper value of
the velocity of light.  As a result we may introduce on  the  path  of
moving the particle some universal time $t=t_0$ like Newtonian time in
classical physics.  Then the velocity of light $c$ will depend on  the
emitter  velocity  according  to the law $c=c_0(1+v^2/{c_0}^2)^{1/2}$.
The  space-time  transformations  retaining  the  invariance  of   the
4-interval $ds^2$ take the form
\end{sloppypar}
\begin{equation}\label{eq.19a}
\begin{array}{l}
\displaystyle
x'=\frac{x-Vt}{\sqrt{1-V^2/c^2}}, \
y'=y,  \ z'=z, \ t'=t,
\end{array}
\end{equation}
where $c'=c(1-Vv_x/c^2)/(1-V^2/c^2)^{1/2}$,  \   $v_x=x/t$.   In   the
Minkowski  space  $M^4(x^0,{\bf  x})$  they  may  be classified as the
transformations of space-time-velocity of light which  belong  to  the
direct  product  of the Lorentz group $L_6$ and the group of the scale
transformations of the velocity of light $\Delta_1(c):  c'=\lambda  c$
($\lambda$  is  the  group  parameter).  The corresponding integral of
action,  Lagrangian,  expressions for 4-momentum and the energy  of  a
particle in electromagnetic field are constructed.  In particular, the
expressions for 3-momentum ${\bf p}$ and energy ${\cal  E}$  take  the
form
\begin{equation}
{\bf p}=m{\bf v}, \ {\cal E}=mc_0c=m{c_0}^2\sqrt{1+v^2/{c_0}^2}.
\end{equation}
As a  result  the theory possesses a number of signs common to both SR
and classical physics.  For example,  as  in  classic  mechanics,  the
particle  mass  $m$  is independent of its velocity $v$.  The particle
energy  ${\cal  E}=mcc_0=m{c_0}^2(1+mv^2/2+\dots)$   in   $v\ll   c_0$
approximation  coincides  with  the  expression  for kinetic energy in
Newton  mechanics  with  an  accuracy  of  the  rest   energy   ${\cal
E}_0=m{c_0}^2$ (as in SR).  The momentum $p$ and the energy ${\cal E}$
are  related  by  the  relationship  coinciding  formally   with   the
relativistic    expression    ${\cal   E}^2-{c_0}^2p^2=m^2{c_0}^4$   -
invariant.  The distinction consists in the fact that here the  energy
and  the  momentum  are  ${\cal  E}=m{c_0}^2(1+v^2/{c_0}^2)^{1/2}$ and
${\bf    p}=m{\bf    v}$,     respectively,     but     not     ${\cal
E}=m{c_0}^2/(1-v^2/{c_0}^2)^{1/2}$       and       ${\bf      p}=m{\bf
v}/(1-v^2/{c_0}^2)^{1/2}$ as in SR.  The equations of  electrodynamics
are nonlinear and take the form
\begin{equation}\label{eq.20}
\begin{array}{ll}
\vspace{1mm}
\displaystyle
\nabla{\rm X}{\bf E}+\frac{1}{c}\frac{\partial{\bf H}}{\partial t}=0, &
\displaystyle
\nabla{\rm X}{\bf H}-\frac{1}{c}\frac{\partial{\bf E}}{\partial t}=
\frac{4\pi}{c}\rho{\bf v},                                           \\
\displaystyle
\nabla\cdot{\bf E}=4\pi\rho, & \nabla\cdot{\bf H}=0,                 \\
\end{array}
\end{equation}
\begin{equation}
\begin{array}{ll}
\displaystyle
\frac{d{\bf p}}{dt}=
\frac{c}{c_0}e{\rm E}+\frac{e}{c_0}{\bf v}{\rm x}{\bf H},            &
\displaystyle
\frac{dc}{dt}=\frac{e}{c_0}{\bf v}\cdot{\bf E},
\end{array}
\end{equation}
where $\nabla c=0$. The equations engage through the velocity of light
$c$.  In $v^2/{c_0}^2\ll 1$ approximation the equations  (\ref{eq.20})
go   into   the   standard   equations  of  classical  electrodynamics
(\ref{eq.4}) with $c=c_0$.

\begin{sloppypar}
Many experiments hitherto interpreted in terms of SR, may be explained
within  the  framework  of  the constructed theory.  For example,  the
experiments of Michelson and Fizeau,  the  aberration  of  light,  the
appearance  of atmospheric $\mu$-mesons near the Earth's surface,  the
Doppler-effect,  a number  of  the  well-known  experiments  to  prove
independence  of  the  velocity  of  light  from the velocity of light
source,  the decay  of  unstable  particles,  the  generation  of  new
particles  from nuclear reactions,  the Compton-effect,  photo-effect.
Let us consider some of them.
\end{sloppypar}

{\it The negative result of the Michelson experiment} for an  observer
with  a  terrestrial source of light (the speed of light $c_0$) may be
explained by the isotropy of space.  Since the speed of light $c_0$ is
the same in all directions, a shift of the interference pattern should
be absent with the interferometer's rotation. Analogously, in the case
of  an  extraterrestrial  source  (the  star  moving  inertialy with a
velocity $v$ relatively to the Earth),  the velocity of light from the
star  $c=c_0(1+v^2/{c_0}^2)^{1/2}$ does not depend on the direction of
its propagation and is thus the same for an observer on the Earth.  As
a   result,   the   interference   pattern   does   not   change  with
the interferometer's rotation.

\begin{sloppypar}
{\it A  free particle with the mass $m$ will move faster than light} at
the velocity $v=c_0[({\cal  E}/{\cal  E}_0)^2-  1]^{1/2}>c_0$  if  the
particle energy satisfies the inequality ${\cal E}>2^{1/2}{\cal E}_0$,
where ${\cal  E}_0=m{c_0}^2$  is  the  particle  rest  energy.  As  an
illustration,  for  the  electron faster-than-light motion begins with
the energy ${\cal E}\sim 723$ keV.  The $1$ MeV-electron  velocity  is
$\sim  1.68c_0$;  the  $1$  GeV-electron  velocity  is $\sim 2000c_0$.
\end{sloppypar}

\begin{sloppypar}
{\it The  angle  of  aberration}  of  the  quasar  Q1158+4635 with the
parameter of red shift $z_{\lambda}=4.73$ (Carswell and Hewett,  1990)
is $\alpha\sim 11.6$ arc seconds;  $z_{\omega}\sim 2.23$, the velocity
of light is $c=c_0(1+z_{\lambda})/(1+z_{\omega})\sim 1.77c_0$  instead
of $\alpha\sim20.5$ arc seconds and $z_{\lambda}=z_{\omega}$,  $c=c_0$
in SR.
\end{sloppypar}

\begin{sloppypar}
{\it The velocity of atmospheric $\mu$-mesons} at the Earth's  surface
is $\sim 100c_0$, the energy is $\sim 10.6$ GeV.
\end{sloppypar}

{\it Independence  of  the  velocity  of light} from the velocity of a
source of light was considered in rather numerous experiments.  As  an
example,  we  turn  to the Filippas and Fox experiment (1964) in which
the    velocities    of     $\gamma$-quanta     from     the     decay
$\pi^0\to\gamma+\gamma$ of fast $\pi^0$-mesons were compared.  Because
of independence of the velocity $c=c_0(1+v^2/{c_0}^2)^{1/2}$ from  the
direction of the $\gamma$-emission, the velocities of $\gamma$-quanta,
as in SR,  should be the same.  It means a  negative  result,  in  the
standard  sense,  of  this  experiment and other experiments of such a
type (Bonch-Bruevich and Molchanov 1956, Sadeh 1963).

\begin{sloppypar}
{\it The    velocity   of   Compton   forward -  scattered   electron}
$v=c_0(\hbar\omega/{\cal   E}_0)[1-   {\cal   E}_0/(2\hbar\omega+{\cal
E}_0)]$ exceeds  the speed of light $c_0$ if the energy of an incident
gamma quantum $\hbar\omega$  is  more  than  $698$  keV.  The  angular
distribution of scattered gamma quanta coincides with the one obtained
from SR. \end{sloppypar}

\begin{sloppypar}
{\it The velocity of photoelectron}
$v=c_0[((\hbar\omega+m_e{c_0}^2-U)/m_e{c_0}^2)^2-1]^{1/2}$ exceeds the
speed of light $c_0$ if the photon  energy  satisfies  the  inequality
$\hbar\omega>211$ keV  +  $U$,  where  $U$  is  the ionization energy,
$m_e{c_0}^2$ is the electron rest energy.

In sum,   the   $L_6{\rm   X}\Delta_1(c)$-invariant  theory  has  been
constructed.  The  faster-than-light  motions  are  possible  in  this
theory. According to the construction, the proposed theory is close to
the SR theory and based (as  SR)  on  the  principles  of  a  symmetry
approach. It has been found that this theory permits one to interpret
a set of the well-known experiments.  In the case of the Michelson and
Fizeau experiments the results of the interpretation coincide with the
accepted results.  On the  other  hand,  there  is  a  number  of  the
predictions which differ from SR predictions.  These are, for example,
faster-than-light   motions    of    nuclear    reactions    products,
faster-than-light motions  in  astrophysics.  These predictions may be
the subject of  experimental  investigation.  The  postulation  $c'=c$
leads to SR.
\end{sloppypar}
\bigskip

\centerline{\bf 10.REFERENCES}
\begin{center}
\begin{tabular}{ll@{\hspace{4cm}}}
\hspace{-3mm}[1] &\hspace{-3mm} 
\begin{sloppypar}G.A. Kotel'nikov. {\bf New Symmetries in Elect-}
\end{sloppypar}                                                       \\
{}& \hspace{-3mm}\begin{sloppypar}{\bf rodynamics and Quantum Theory}. 
Synopsis  \end{sloppypar}                                         \\
{}& \hspace{-3mm}\begin{sloppypar}of thesis for a Doctor's degree, \ RRC \
Kurchatov \end{sloppypar}                                       \\
{}& \hspace{-3mm}\begin{sloppypar}Institute, Moscow, 1999, 41 p.\end{sloppypar}
\end{tabular}
\begin{tabular}{ll@{\hspace{4cm}}}
\hspace{-3mm}[2]&\hspace{-3mm} 
\begin{sloppypar}G.A. Kotel'nikov. {\bf On the Faster-Than-Light}
\end{sloppypar}                                                       \\
{}&\hspace{-3mm}\begin{sloppypar}{\bf Motions in Electrodynamics}.
Proc. \ XII \ Int.  \end{sloppypar}                               \\
{}&\hspace{-3mm}\begin{sloppypar}Conf. on Selected Problems of \ Modern \ Physics,
\end{sloppypar}                                         \\
{}&\hspace{-3mm}\begin{sloppypar}Dubna, June 8-11, 2003, D1, 2-2003-219, 
143-147; \end{sloppypar} \\
{}&\hspace{-3mm}physics/0311041. 
\end{tabular}
\end{center}
\end{document}